\documentclass[aip,jmp,preprint]{revtex4-1}
\usepackage{graphicx}
\usepackage{tabularx}
\usepackage{caption} 
\usepackage{color}
\newcommand{\SBH}{SBH}
\newcommand{\SBHS}{SBHs}
\newcommand{\apj}{\it Astrophys. J.}

\newcommand{\apjl}{\it Astrophys. J. Lett.}
\newcommand{\aap}{\it Astron. Astrophys.}

\newcommand{\mnras}{\it Mon. Not. R. Astron. Soc.}
\newcommand{\aj}{\it Astron. J.}
\newcommand{\araa}{\it Annu. Rev. Astron. Astr.}

\newcommand{\nat}{\it Nature}
\newcommand{\prd}{\it Phys. Rev. D}

\begin{document}

\title{A candidate sub-parsec binary black hole in the Seyfert galaxy NGC\,7674}
\author{P. Kharb}
\affiliation{National Centre for Radio Astrophysics - Tata Institute of Fundamental Research, Pune University Campus, Post Bag 3, Ganeshkhind, Pune 411007, India}
\author{D. V. Lal}
\affiliation{National Centre for Radio Astrophysics - Tata Institute of Fundamental Research, Pune University Campus, Post Bag 3, Ganeshkhind, Pune 411007, India}
\author{D. Merritt}
\affiliation{School of Physics and Astronomy and Center for Computational Relativity and Gravitation, Rochester Institute of Technology, 54 Lomb Memorial Drive, Rochester, NY 14623, USA}

\begin{abstract}
{\bf The existence of binary supermassive black holes (\SBHS) is predicted by models of hierarchical galaxy formation. To date, only a single binary \SBH\ has been imaged, at a projected separation of 7.3 parsecs. Here we report the detection of a candidate dual \SBH\ with projected separation of 0.35 pc in the gas-rich interacting spiral galaxy NGC\,7674 (Mrk\,533). This peculiar Seyfert galaxy possesses a $\sim$0.7~kpc Z-shaped radio jet; the leading model for the formation of such sources postulates the presence of an uncoalesced binary \SBH\ created during the infall of a satellite galaxy. Using very long baseline interferometry (VLBI), we imaged the central region of Mrk\,533 at radio frequencies of 2, 5, 8 and 15~GHz. Two, {possibly inverted-spectrum} radio cores were detected at 15~GHz only; the $8-15$~GHz spectral indices of the two cores are $\ge-0.33$ and $\ge-0.38$ ($\pm 30\%$), consistent with accreting \SBHS. We derive a jet speed $\sim0.28c$ from multi-epoch parsec-scale data of the hotspot region, and a source age $\ge8.2\times10^3$~yrs.}
\end{abstract}
\maketitle

Mrk\,533 (a.k.a. NGC\,7674) is a nearby {($z=0.028924$, luminosity distance $D_L\approx116$ Mpc)} luminous infrared galaxy (LIRG). LIRGs have infrared ($8-1000~\mu$m) luminosities in the range $10^{11}-10^{12}$~L$_\odot$ and are  interacting systems.\citep{sanders96} Mrk\,533 is the brightest member of the Hickson~96 compact galaxy group consisting of four galaxies.\citep{verdes1997} Clear tidal tails are observed in the Digitized Sky Survey\cite{Lasker90} images and it appears that a merger with a small neighboring galaxy is in progress. Mrk\,533 has been classified as a Seyfert type 2 active galactic nucleus (AGN); i.e., its spectra exhibit narrow emission lines of velocity width $\sim100-1000$~km~s$^{-1}$. AGN are believed to result from mass accretion on to supermassive black holes (SBHs) with masses $10^6$~M$_\odot \lesssim$ M$_{\bullet} \lesssim10^9$~M$_\odot$; the emission lines in the spectra are believed to be emitted from gas clouds orbiting the SBH$-$accretion disk system in regions of sub-parsec size (the broad emission line region, BLR) or at distances of hundreds to thousands of parsecs (the narrow emission line region, NLR). While Mrk\,533 is a nearly face-on spiral galaxy, its optical spectrum shows a broad H$\beta$ line in polarized emission,\citep{Miller90} implying that the view of the central AGN, specifically the BLR is obscured, presumably by a dusty torus. 

Hierarchical models of galaxy formation predict that spiral galaxies merge to form elliptical galaxies, leaving two or more \SBHS\ in the center of the merged elliptical.\cite{Begelman80} The host galaxies of nearby Seyfert galaxies, which are spiral in nature, are expected to evolve via minor mergers.\citep{Aguerri01} Since small galaxies are believed less likely to contain massive \SBHS\ than massive galaxies,\citep{Kormendy13} it is not a priori clear whether Seyfert galaxies should ever host binary \SBHS. Nevertheless we report here the detection of sub-parsec-scale dual radio cores which we interpret as a binary \SBH\ in the interacting Seyfert galaxy Mrk\,533. 

Mrk\,533 has radio jets on the $\sim$700 parsec scale.\citep{Momjian03} In this paper, we present multi-frequency parsec-scale data on Mrk\,533 acquired with the Very Long Baseline Array (VLBA), showing the presence of two compact radio components. We adopt the following cosmological parameters: H$_0=73$~km~s$^{-1}$~Mpc$^{-1}$, $\Omega_\mathrm{matter}=0.27$ and $\Omega_\mathrm{vacuum}=0.73$. At the distance of Mrk\,533, 1 arcsecond corresponds to 0.533 kpc. The radio spectral index, $\alpha$, is defined such that flux density, $F_\nu$, at frequency $\nu$ is $F_\nu\propto\nu^\alpha$.

\section{Results}
Momjian et al.\cite{Momjian03} argued that the AGN in Mrk\,533 was located along a line connecting the two strongest sources, which they labeled ``C" and ``W", and that those two, kiloparsec-scale features were radio lobes or hot spots created by the AGN. However, no radio core was detected by them in their very long baseline interferometry (VLBI) data at 1.4~GHz. We looked at recently acquired Expanded Very Large Array (EVLA) A-array 15~GHz data, at a resolution of $\sim$100~milliarcseconds (mas), in the NRAO archive. We found that the radio core that was missing in the 1.4~GHz data of Momjian et al.\cite{Momjian03} is now clearly detected at 15~GHz (Figure~\ref{fig1}), implying that it has an inverted spectral index (see the Methods section). 

In our new $\sim$1~mas resolution VLBI image at 2~GHz, we detect the south-east and north-west hotspots (Figure~\ref{fig2}), corresponding to the C and W components of Momjian et al.\cite{Momjian03}; for ease of comparison, we will continue to refer to them as the C and W components, respectively. In addition, we detect two radio cores at 15~GHz along the connecting line between C and W, and coincident with the $\sim$100~mas radio core (Figures~\ref{fig3}, \ref{fig2}, \ref{fig1}). The eastern and western cores (C1, C2) are detected at the 4 to 5$\sigma$ level. Their positions are: right ascension (RA) 23h 27m 56.70060s, declination (DEC) 08$^\circ$ 46$^\prime$ 44.21208$^{\prime\prime}$, and RA 23h 27m 56.70055s, DEC 08$^\circ$ 46$^\prime$ 44.21213$^{\prime\prime}$. The projected separation between the two cores is 0.65 mas, or 0.35~parsec at the distance of Mrk\,533. They are $\approx14.0\pm0.4$~mas and $\approx6.0\pm0.4$~mas away in RA and DEC, respectively, from the peak position of the {EVLA core region}, whose size is 100~mas $\times$ 20~mas. The integrated flux density of the EVLA core region (= 1.88 $\pm~0.02$ mJy) is nearly equal to the sum of the integrated flux densities of the two VLBI cores (= 1.8 $\pm~0.5$ mJy), implying that, going from $\sim$100~mas to $\sim$1~mas scales, nearly all of the compact core emission is accounted for. 

Spectral index images between $2-5$~GHz and $5-8$~GHz indicate that components C and W have steep radio spectra, consistent with optically thin synchrotron jets. For the cores detected only at 15~GHz, the $8-15$ GHz spectral index, derived using four times the {\it r.m.s.} noise in the 8~GHz image, is $\ge-0.33$ and $\ge-0.38$ for the eastern and western core components, respectively. The errors in these spectral index values could be 30\% or larger (see the Methods section). Since the cores are not detected at frequencies equal to or lower than 8~GHz, they have inverted spectra at these frequencies, as are typically observed {in AGN cores}.\citep{Barvainis96,Kharb14} This implies that the synchrotron self-absorption (SSA) turnover may have occurred around or below 15~GHz in this source, similar to GHz peaked spectrum (GPS) sources\cite{Hancock10} (see Figure~\ref{fig5}). The inverted spectra could also be a consequence of free-free absorption (FFA), as suggested by Momjian et al.\cite{Momjian03} While there is insufficient spectral index information for the cores in the self-absorbed regions to distinguish between the SSA and FFA mechanisms, their high brightness temperatures (see below) support the SSA mechanism for the inverted spectra.

The brightness temperature, estimated using the relation in Ulvestad et al.\cite{Ulvestad05} for an unresolved component, is $\sim5.8\times10^7$~K for the eastern core. As the minor axis of the western core could not be deconvolved from the beam in {the Gaussian-fitting software} (see the Methods section), we used the beam minor axis to obtain its brightness temperature: it was $>1.6\times10^7$~K assuming an unresolved component. These high brightness temperatures are indicative of non-thermal emission as observed in the unresolved bases of radio jets in AGN. 

While radio supernovae or supernova remnants are known to possess sizes ($\sim0.1-0.5$~pc)\cite{Varenius17} and  brightness temperatures ($\gtrsim10^7$K)\cite{Perez09} similar to the two radio cores, the accurate positional and flux density constraints relative to the EVLA core region make them unlikely candidates for the two cores in Mrk\,533. The implied ``inverted'' spectra of the two compact features do not favour a one-sided core-jet structure or two hotspots of a single radio AGN; compact jet components or hotspots typically have relatively flat spectra. Furthermore, the probability of finding two spurious 5$\sigma$ noise peaks at the positions of the cores is $<0.4\%$ (see  the Methods section). On the other hand, the positional coincidence of the two cores with the 15~GHz EVLA core region; their high brightness temperatures; and the implied, inverted spectral indices, are all consistent with the two cores being two accreting \SBHS. This would make the AGN in Mrk\,533 the binary \SBH\ having the smallest observed separation to date. In fact, the only other parsec-scale binary that has been imaged is in the radio galaxy 4C\,+37.11 (B2 0402+37) which is hosted by an elliptical galaxy; the projected separation is 7.3 pc.\citep{Rodriguez06} A 0.1~pc separation binary was {\it inferred} from the presence of two BLRs in the quasar SDSS~J153636.22+044127.0 by Boroson \& Lauer\cite{Boroson09} (however see \citep{Chornock09,Wrobel09}). 

\subsection{Black Hole Masses}
Woo \& Urry\cite{Woo02} measured a central velocity dispersion of the stars in Mrk\,533 of 144~km~s$^{-1}$; assuming that the \SBH\ in this galaxy respects the $M_\bullet-\sigma$ relation \cite{Ferrarese00} they inferred a \SBH\ mass of $\sim3.63\times10^7$~M$_\odot$. They also estimated the bolometric luminosity of the AGN using flux integration to be $L_\mathrm{bol}=1.41\times10^{45}$~erg~s$^{-1}$. This implies that the source has an Eddington luminosity ($\equiv1.25\times10^{38}(\mathrm{M_\bullet/M_\odot})$~erg~s$^{-1}$) of $\sim4.54\times10^{45}$~erg~s$^{-1}$ and an Eddington ratio (i.e., the ratio of bolometric luminosity to Eddington luminosity) of $\sim$0.3. This would place Mrk\,533 at the high end of Eddington ratio distribution for Seyfert galaxies.\citep{Ho08} 
{While the distance of the radio ``core" from the SBH is still an open question, Hada et al.,\cite{Hada11} found it to be between $14-23$ gravitational radii ($R_s\equiv 2GM_\bullet/c^2$) in the radio galaxy M87. If we assume that the radio core in Mrk\,533 is similarly at a distance of $10R_s$, this translates to a spatial extent of 35 microparsecs (= 7.2 AU). This rules out the possibility that the two radio cores  in Mrk\,533 are the bases of the jet and counterjet from a single AGN.}

The sizes of BLRs in Seyferts range from $10-20$ light days,\cite{Peterson88,Nunez15} making them much smaller than the separation of our putative binary ($\sim420$ light days). If the two SBHs are similar in terms of their masses and emission properties, double-peaked emission lines should have been visible from the individual BLRs around the two SBHs. However, the polarized broad-line spectrum of Miller \& Goodrich\cite{Miller90} shows the broad lines to be relatively narrow ($\sim2000$~km~s$^{-1}$), making it difficult to identify small splits in their peaks; the presence or absence of double peaks cannot be confirmed with those data. Sizes of NLRs are much larger than the binary separation, and the gas making up the two NLRs would be expected to have merged, consistent with the absence of double-peaked narrow emission lines.

A massive binary at the center of a galaxy is deemed ``hard'' if its separation is less than $\sim a_h$, where\citep{Merritt13}
\begin{eqnarray}
a_h = \frac{G\mu}{4\sigma^2} \approx 0.27 \left(1+q\right)^{-1} \left(\frac{M_2}{10^7 M_\odot}\right)
\left(\frac{\sigma}{200\ \mathrm{km}\ \mathrm{s}^{-1}}\right)^{-2} \mathrm{pc}.
\end{eqnarray}
{Here, $q\equiv M_2/M_1$ and $\mu\equiv M_1M_2/(M_1+M_2)$, where $M_1$ and $M_2$ are the binary's component masses.}
Setting $M_1+M_2=3.6\times10^7 M_\odot$ and $\sigma=144$ km s$^{-1}$ yields 
$a_h\approx0.47$ pc assuming $q=1$ ($a_h\approx 0.85$ pc assuming $q=0.1$).
Given our estimated projected separation of $0.35$~pc, it is reasonable to conclude that the binary is roughly at the hard-binary separation. A separation $\sim a_h$ is where a massive binary in a stellar nucleus would be expected to ``stall,'' since at this separation the binary is able to eject from the nucleus stars that interact closely with it, shutting off the gravitational slingshot interactions that would allow further hardening.\cite{Merritt13} 

\subsection{Jet Speed and Source Age}
In order to determine the jet speed in Mrk\,533, we obtained archival VLBI data at 5~GHz from 1998\citep{Lal04} and looked at the C hotspot region to see if any jet knots were moving toward this hotspot (HS). This was found to be the case for knots J1 and J2 (see Figure~\ref{fig6}). We could not use the RA and DEC positions for the knots though, as the 1998 experiment was not phase-referenced. We therefore used the relative separation between knots J1 and J2 in 1998 and then in 2002 (current project). We found a jet speed of $0.28c$, a value consistent with other Seyfert jet speed estimates in the literature.\citep[e.g.,][]{Bicknell98,Ulvestad99} As the jet speeds at jet launching sites are presumably larger than the ones estimated at hotspots, due to the expected deceleration, the kiloparsec-scale jet speed derived above is likely to be a lower limit. Using the jet speed of $0.28c$ and jet extent of 0.7~kpc, we derive a lower limit for the source age to be $8.2\times10^3$~yrs. This estimate is consistent with the Mrk\,533 being a GPS source, which are typically young. Using the European VLBI Network (EVN) at 1.5~GHz (resolution $\sim$ 30~mas), Middelberg et al.\cite{Middelberg04} estimated the hotspot advance speed in Mrk\,533 from the relative separation of components C and W between the years 1985 and 1999, to be $0.92c$. Even though the jet speed estimated above ($0.28c$) is a lower limit, it can be used in conjunction with the hotspot advance speed of $0.92c$ to derive the ratio of the jet density to the ambient density $\eta\equiv\rho_{jet}/\rho_{amb}$ using simple one-dimensional momentum conservation arguments.\cite{deYoung02} The resultant $\eta=2$ implies a `heavy' jet. Alternately, the simple momentum conservation arguments may not fully apply if the jet propagation direction is changing. Our 1998 image also showed a jet component beyond the hotspot C: this knot (J0 in Figure~\ref{fig6}) which should be moving away from the hotspot, is not observed in 2002. Presumably it has moved, expanded and faded in intensity.  

\subsection{The Z-Shaped Morphology}

That Mrk\,533 contains an uncoalesced, binary \SBH\ is also suggested by the Z-shaped morphology of the radio source on kiloparsec scales.
Momjian et al.\cite{Momjian03}, Gopal-Krishna et al.\cite{Gopal03} and Zier et al.\cite{Zier05}
have argued that radio sources associated with a galactic merger, in which both galaxies contain a \SBH, should evolve along a Z$\rightarrow$X morphological sequence; transition to an X-shaped source occurs after the two \SBHS\ coalesce and undergo a spin-flip.\cite{Merritt02} Prior to coalescence, the radio jets associated with the active \SBH\ are influenced by hydrodynamic forces, either from gas  brought in with the secondary galaxy, or ambient gas that is perturbed by it; this gas circulates within the larger galaxy and its angular momentum will typically be mis-aligned with respect to the original jet direction, i.e., the spin axis of the active \SBH. Pressure forces from the rotating gas can push the jets apart, so that they come to constitute the two parallel arms of the Z.\cite{footnote} If the binary \SBH\ manages to overcome the ``final-parsec problem''\cite{Milosavljevic03} and coalesce, a spin-flip can occur when the larger \SBH\ absorbs the smaller \SBH. The accretion disk, which determines the jet direction, will then re-orient with respect to the new \SBH\ spin direction via the Bardeen-Petterson effect.\cite{Bardeen75} If this reorientation is sufficiently rapid, the result will be an X-shaped source, with the two arms of the Z now comprising the passive lobes, or ``wings,'' of the X. 

In this picture, the NE and SW lobes of Mrk 533 would originally have been co-linear with each other and with the spin axis of the larger \SBH, before being displaced by hydrodynamic forces. We have argued based on the VLBI images that the {\it current} jet direction is {\it not} parallel to the two lobes. Some mechanism must therefore have acted to rotate the small-scale jet -- which appears to be generated in the eastern core -- in an anti-clockwise direction compared with its original orientation. This would be consistent with the hotspots having steep radio spectra ($<-1$, see Figure~\ref{fig4}; also Middelberg et al.\cite{Middelberg04}), since the working surface for the terminal jet shock is constantly moving, albeit slowly.

One mechanism capable of changing the spin axis of the \SBH\ is ``geodetic" precession due to spin-orbit (Lense-Thirring) torques from the orbiting \SBH. The timescale for the spin axis of the primary \SBH\ to precess due to torques from the second \SBH\ is\cite{Merritt13}
\begin{eqnarray}
T_\mathrm{precess} &\approx& \frac{1-e^2}{2q} \left(\frac{a}{R_M}\right)^{3/2} 
\frac{a}{c} \nonumber \\
&\approx& 3\times 10^7 \mathrm{yr}\;  q^{-1} 
\left(\frac{a}{0.35 \mathrm{pc}}\right)^{5/2}
\left(\frac{M_\bullet}{5\times 10^7 M_\odot}\right)^{-3/2} 
\label{Equation:Tprecess}
\end{eqnarray}
where $M_\bullet$ is the mass of the precessing \SBH, $q\equiv M_2/M_\bullet$ is the mass ratio, $R_M\equiv GM_\bullet/c^2$ is the gravitational radius of the precessing \SBH, $a$ is the semi-major axis of the binary orbit  and $e$ is the orbital eccentricity. Due to the strong $a-$ dependence of $T_\mathrm{precess}$, precession would be expected to ``turn on'' suddenly as the two \SBHS\ come together. 

\subsection{Gravitational Wave Signal}
At a distance of $\sim116$~Mpc, the binary \SBH\ in Mrk\,533 would be a factor of two nearer to the Earth than the binary imaged by Rodriguez et al.,\cite{Rodriguez06} and much nearer than the putative close binaries in the blazars OJ\,287 \cite{Valtonen07} or PG\,1302$-$102,\cite{Graham15} both of which are at redshifts $z\approx0.3$. In spite of its nearness, the wide separation ($\sim1$ pc) inferred for our binary implies an orbital frequency that is too low for effective detection by existing or planned gravitational wave detectors. A Keplerian binary has inverse period
\begin{equation}
P^{-1} \approx 3\times 10^{-12} \left(\frac{M_{1} + M_{2}}{10^8M_\odot}\right)^{1/2} \left(\frac{a}{\mathrm{pc}}\right)^{-3/2}
\mathrm{Hz} .
\end{equation}
The peak sensitivity of a pulsar timing array (PTA) occurs at a frequency that is roughly the inverse of the time over which pulsar timing data has been collected; that time is now roughly 10 yr corresponding to a frequency of $\sim3\times 10^{-9}$ Hz~~\cite{Cordes13}. Given current sensitivities, PTAs can detect individual sources at $\sim10^2$ pc distances only if they are both massive ($\gtrsim10^9 M_\odot$) and compact (semi-major axis $a\lesssim10^{-2}$ pc)\cite{Huerta15,Babak16} meaning that the putative binary in Mrk\,533 would likely be detectable only as one among thousands that contribute to the stochastic GW background.\cite{Rasskazov16} Prospects for detection would be even worse in the case of a space-based interferometer like the proposed eLISA, which would be sensitive to $\sim10^8 M_\odot$ binaries only with frequencies above $\sim10^{-5}$ Hz~~\cite{Amaro13}.

The corresponding author is P. Kharb. Requests for materials can be sent to kharb@ncra.tifr.res.in. 

\acknowledgments
DM was supported by the National Science Foundation under grant no. AST 1211602 and by the National Aeronautics and Space Administration under grant no. NNX13AG92G. The National Radio Astronomy Observatory is a facility of the National Science Foundation operated under cooperative agreement by Associated Universities, Inc.

\section*{Author Contributions} 
P.K. reduced the 2002 VLBA data, coordinated the research and wrote the manuscript. D.V.L. led the VLBA proposals for the 2002 and the 1998 data, reduced the 1998 data, and provided feedback on the manuscript. D.M. provided the theoretical framework for the project. P.K. and D.V.L. thank Denise C. Gabuzda for help in obtaining the 1998 and 2002 VLBA data and providing expert knowledge on VLBI data reduction.

\begin{figure}
\centerline{
\includegraphics[width=18cm]{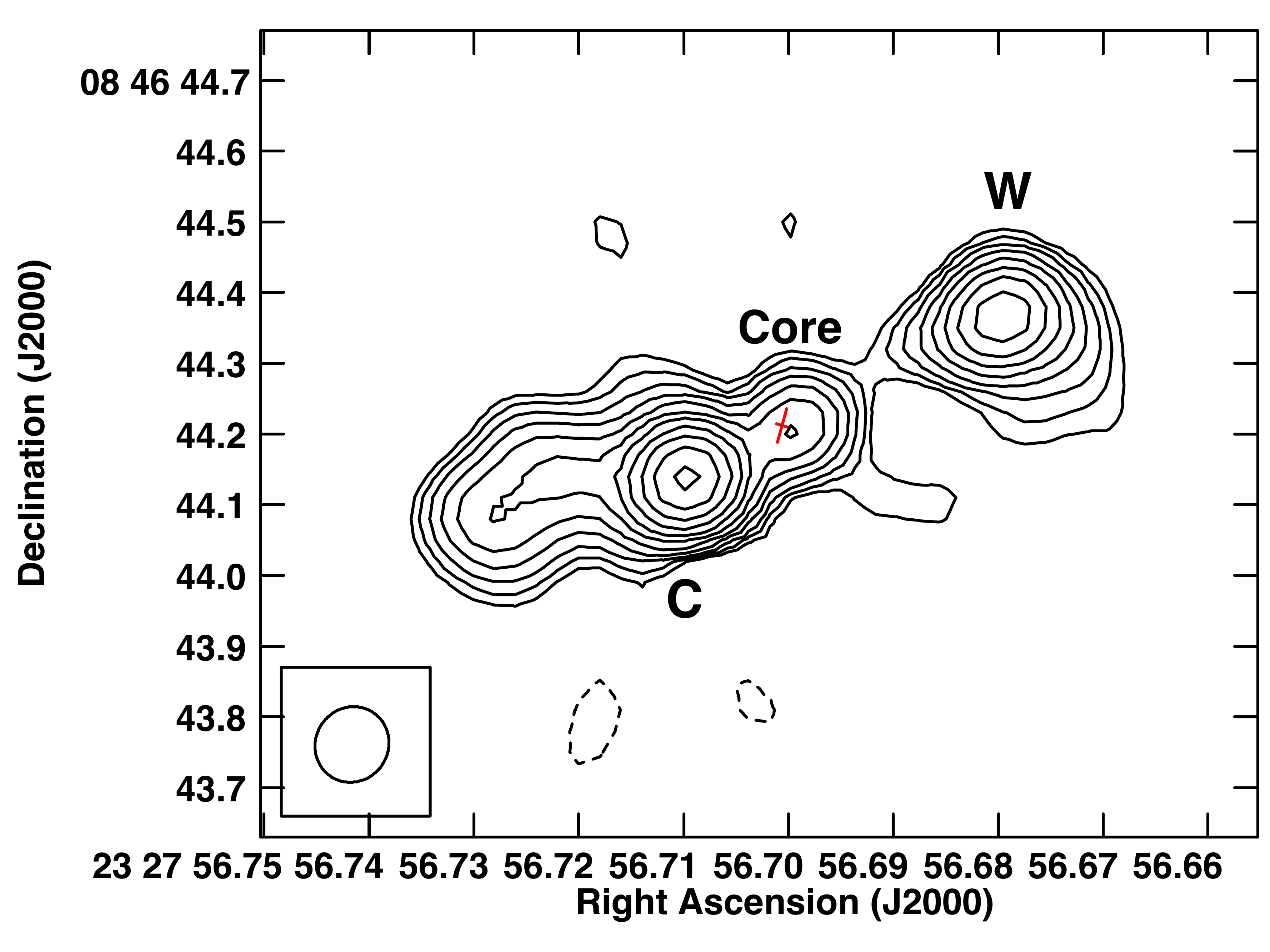}}
\caption
{{\bf 15~GHz contour image of Mrk\,533 from the EVLA in the A-array configuration showing the C and W hotspots as well as a $\sim$ 100~mas radio core detected for the first time in this galaxy.} The beam plotted in the bottom left corner is of size 108~mas~$\times~103$~mas at PA$=-32^\circ$. The contours are in percentage of the peak surface brightness ($=5.72$~mJy~beam$^{-1}$) and increase in steps of $\sqrt 2$, with the lowest contour level being $\pm2.8\%$. The {\it r.m.s.} noise in the image is $\sim~1.0\times10^{-5}$~Jy~beam$^{-1}$. The red cross denotes the position of the eastern radio core, with the VLBI synthesized beam multiplied by a factor of 25. Dashes represent negative contours in this and other contour images. Right Ascension in all contour images is in hours\,:\,minutes\,:\,seconds, Declination in degrees\,:\,arcminutes\,:\,arcseconds.}
\label{fig1}
\end{figure}

\begin{figure}
\centerline{
\includegraphics[width=18cm]{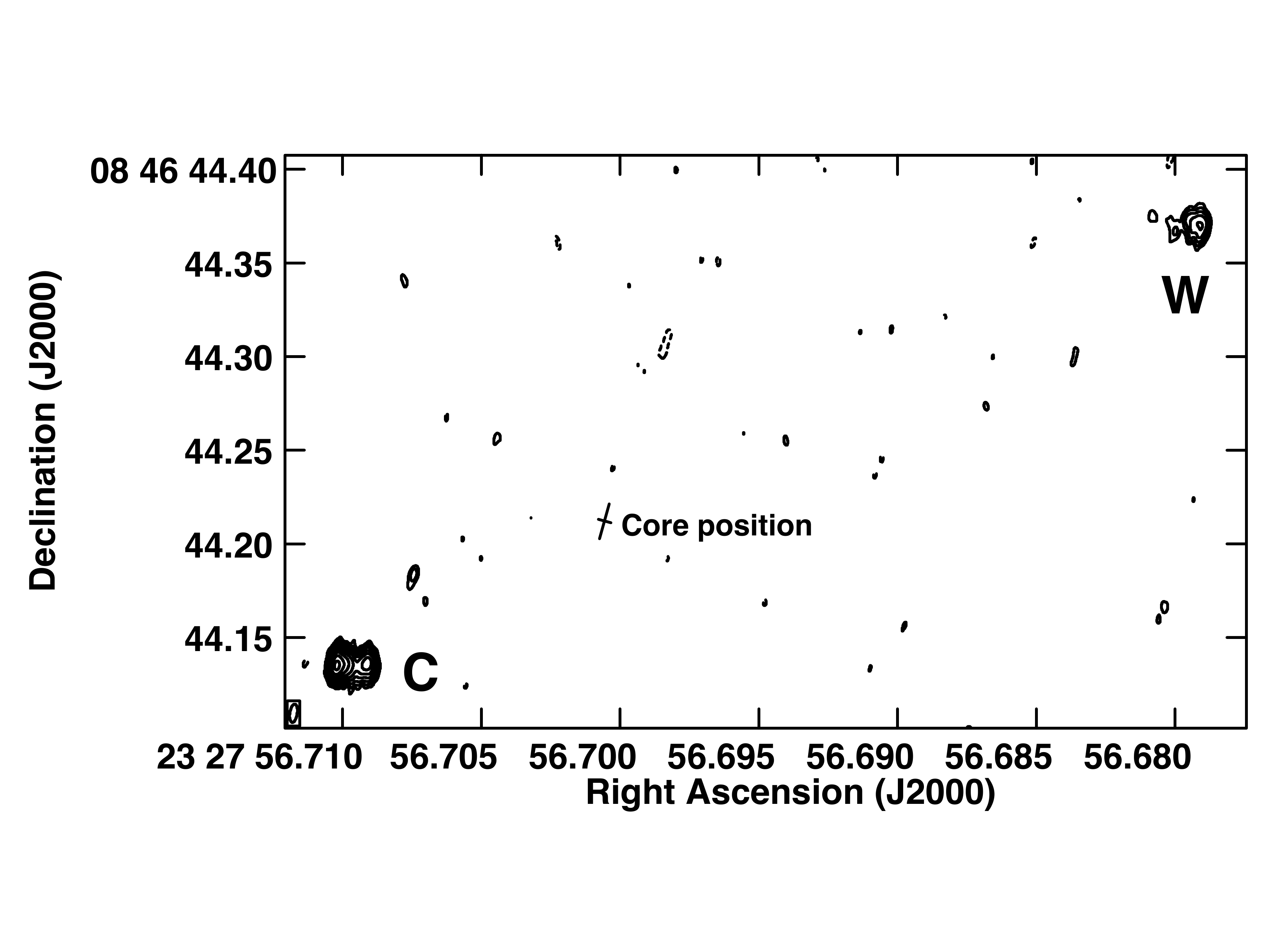}}
\caption
{{\bf VLBA contour image at 2~GHz showing the South-East (C) and North-West (W) hotspots, along with the position of the cores detected at 15 GHz as a cross.} This image was obtained directly from phase-reference visibilities and is not self-calibrated. The cross denotes the position of the eastern radio core, with the 15~GHz beam multiplied by a factor of ten. The 2~GHz beam is shown in the lower left corner; it has a size of 0.98~mas~$\times~0.39$~mas at PA=$-8^\circ$. The contours are in percentage of the peak surface brightness ($=8.78$~mJy~beam$^{-1}$) and increase in steps of $\sqrt 2$, with the lowest contour level being $\pm5.6\%$. The {\it r.m.s.} noise in the image is $\sim~1.0\times10^{-4}$~Jy~beam$^{-1}$. }
\label{fig2}
\end{figure}

\begin{figure}
\centerline{
\includegraphics[width=18cm]{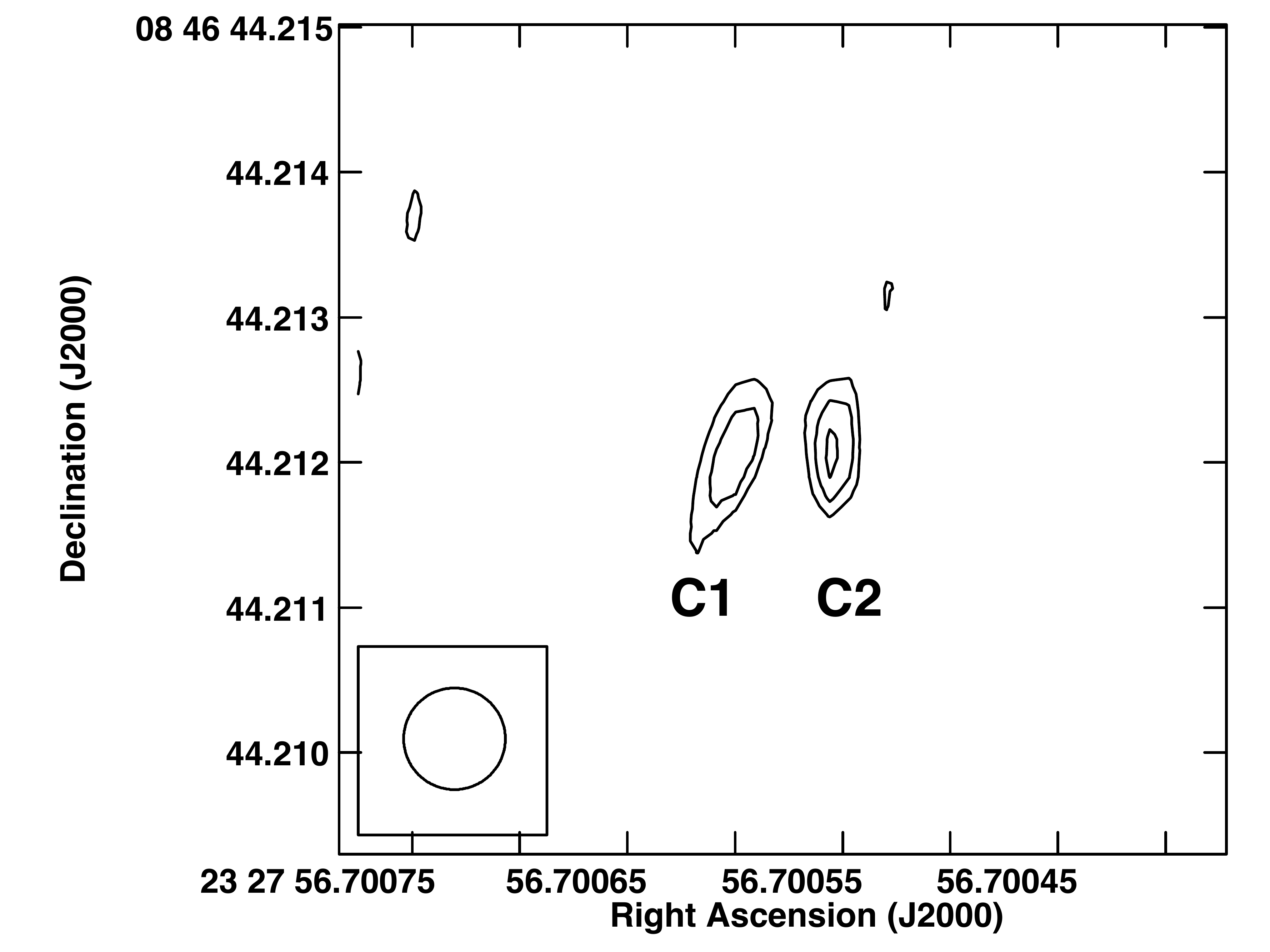}}
\caption
{{\bf VLBI images showing the two cores (C1, C2) detected at 15~GHz.} The peak intensity is 0.74 and 0.76 mJy~beam$^{-1}$ for C1 and C2, respectively (see Table~1). The contour levels are $\pm$0.655, 0.748, 0.855 mJy~beam$^{-1}$, while the beam plotted in the bottom left corner is of size 0.7~mas~$\times~0.7$ mas. The {\it r.m.s.} noise in the image is $\sim~2.0\times10^{-4}$~Jy~beam$^{-1}$. }
\label{fig3}
\end{figure}

\begin{figure}
\centerline{
\includegraphics[width=22cm]{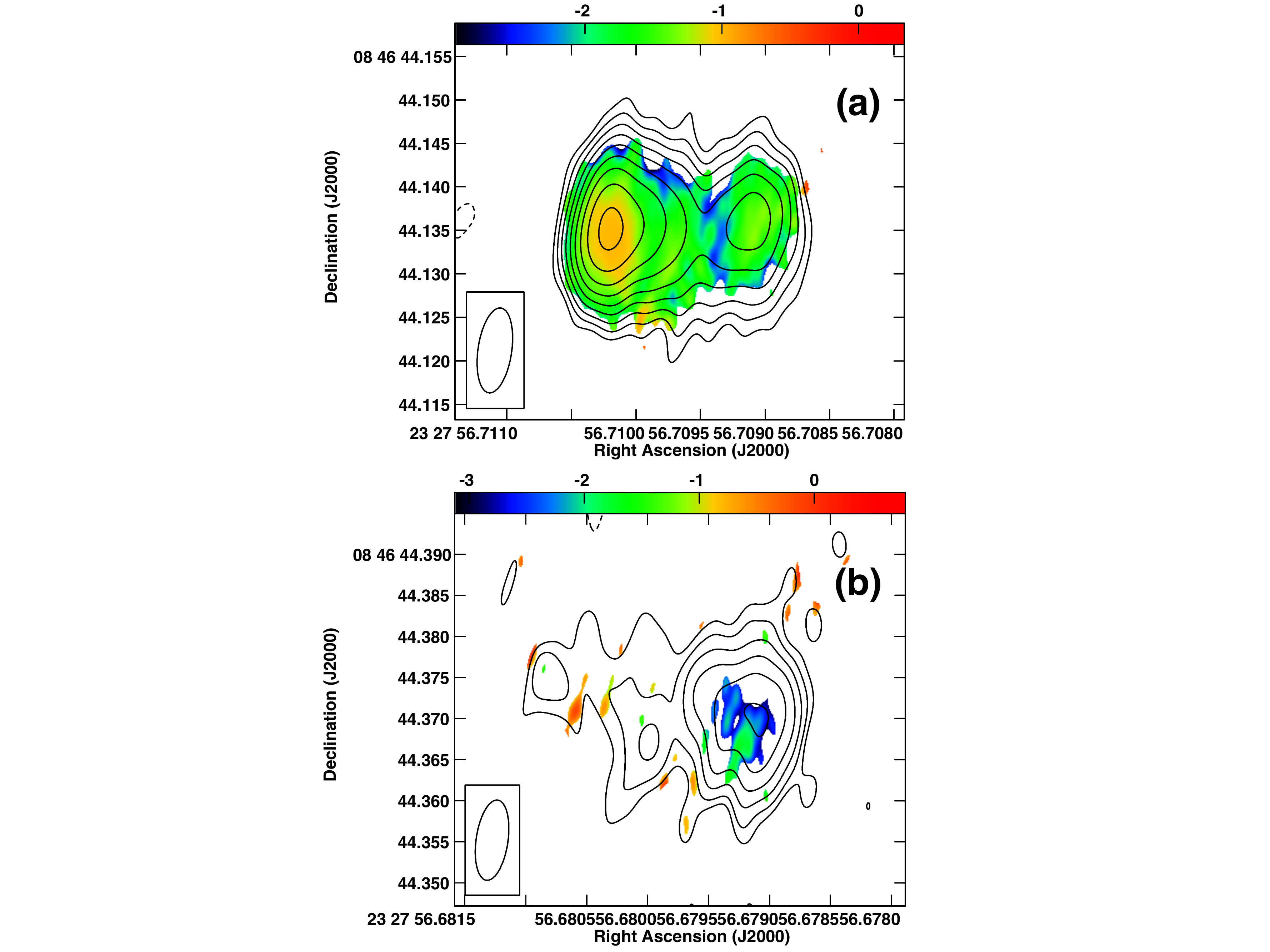}}
\caption
{{\bf A $2-5$~GHz spectral index image of the two hotspots. a, Southeast hotspot. b, Northwest hotspot.} The 2~GHz contours are in percentage of the peak surface brightness ($= 8.78$~mJy~beam$^{-1}$) and increase in steps of $\sqrt 2$, with the lowest contour level being (a) $\pm5.6\%$ and (b) $\pm4.0\%$, respectively. The {\it r.m.s.} noise in the 2 GHz image is $\sim~1.0\times10^{-4}$~Jy~beam$^{-1}$. The beam plotted in the bottom left corner is of size 9.8~mas~$\times~3.9$~mas at PA$=-8^\circ$. 
The color bar indicates the range of spectral index values. Typical uncertainty in the spectral index is $\sim~20\%$.}
\label{fig4}
\end{figure}

\begin{figure}
\centerline{
\includegraphics[width=18cm]{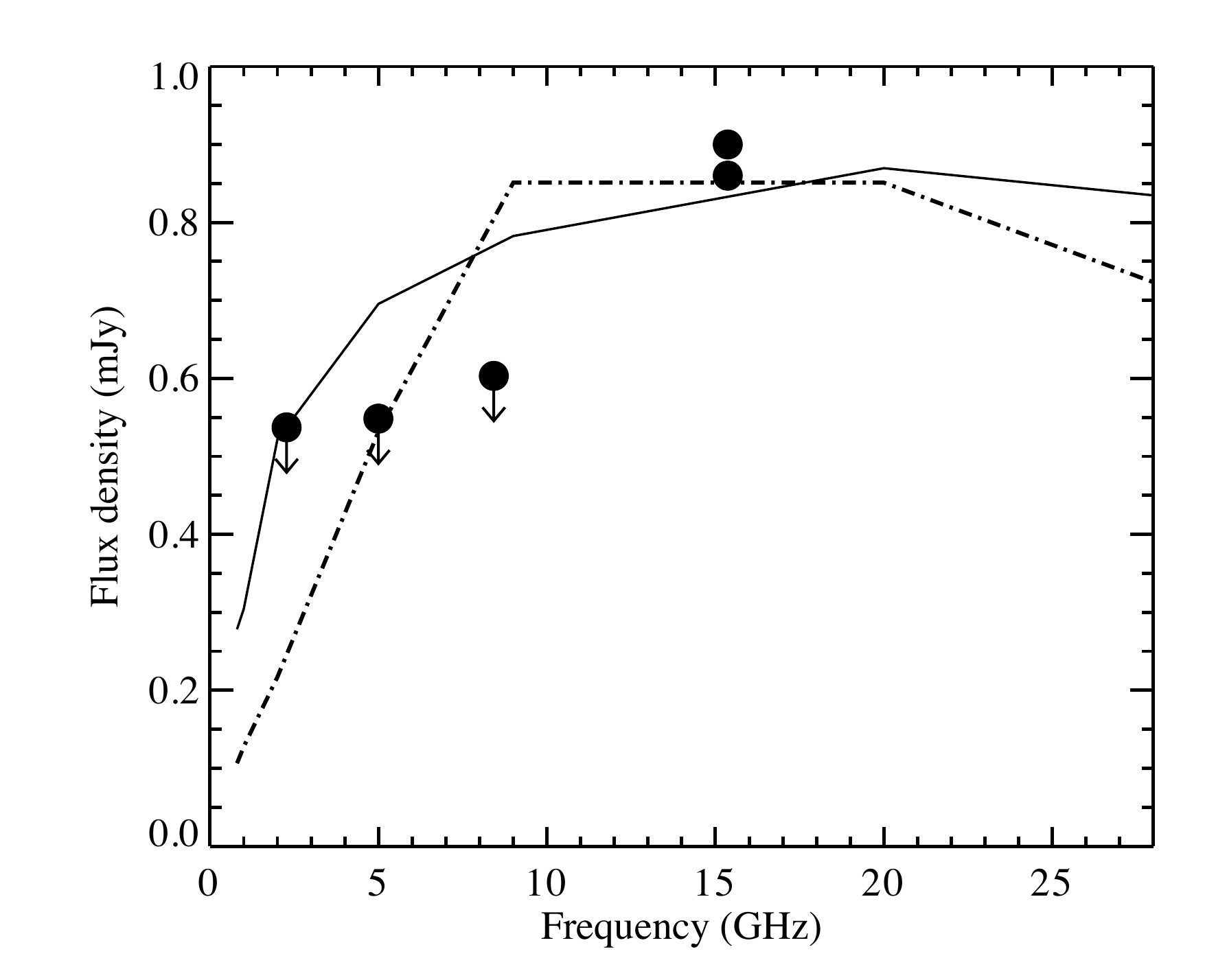}}
\caption{{\bf Radio spectrum of the two VLBI cores in Mrk\,533, which are only detected at 15 GHz}. Data points or the 4$\sigma$ upper limits are shown as filled circles. The total flux densities of the cores have an error of $\approx$35\%. The lines show the spectra/spectral-fits from the GPS sources J005427$-$341949 (solid line) and J181225$-$712006 (dot-dashed line) from Hancock et al.\cite{Hancock10}.}
\label{fig5}
\end{figure}

\begin{figure}
\centerline{
\includegraphics[width=22cm]{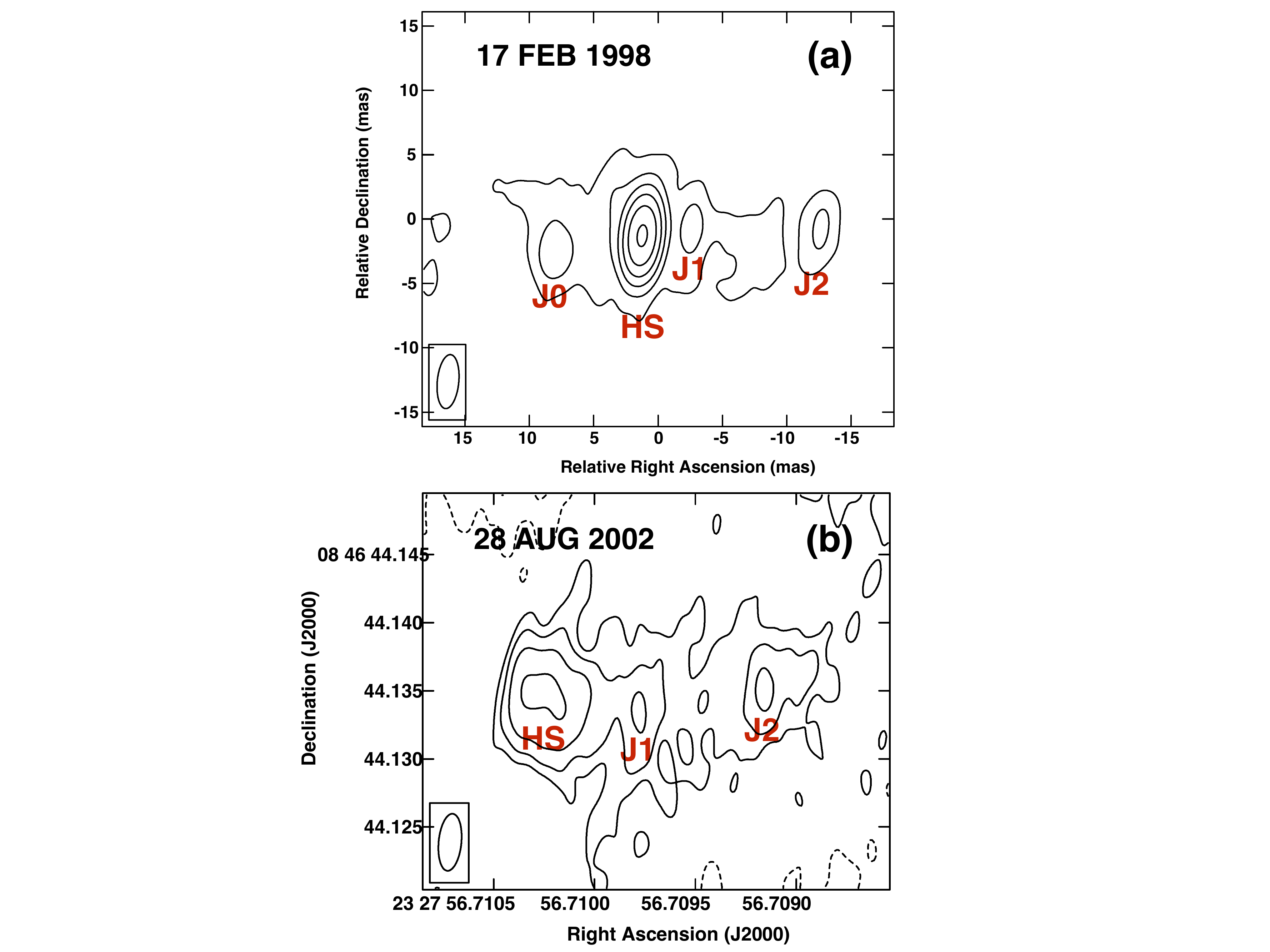}}
\caption
{{\bf 5~GHz contour images of the SE hotspot region of Mrk\,533 from 1998 and 2002.} The hotspot and various jet knots are labelled as HS, J0, J1 and J2. A jet speed of $0.28c$ was derived from the relative separation between jet knots J1 and J2. The contour levels are in percentage of peak surface brightness and increase in steps of 2: the peak surface brightness and lowest contour levels are (a) 8.65~mJy~beam$^{-1}$, $\pm2.8\%$, and (b) 1.79~mJy~beam$^{-1}$, $\pm11.3\%$, respectively. The {\it r.m.s.} noise in the images is (a) $\sim~1.1\times10^{-4}$~Jy~beam$^{-1}$ and (b) $\sim~1.4\times10^{-4}$~Jy~beam$^{-1}$. The beam plotted in the bottom left corner is of size 4.2~mas~$\times~1.7$~mas at PA$=-6^\circ$. }
\label{fig6}
\end{figure}

\begin{table}
\caption{Source Parameters}
\begin{tabular}{cccccc}
\hline\hline
{Comp} & {Freq } & {$\mathrm{I_{peak}}$} & {F$_\mathrm{total}$ } & {$a\times b$} & {$a\times b$} \\
                      & (GHz) & (mJy~beam$^{-1}$) & (mJy) & (mas$\times$mas) & (pc$\times$pc)  \\ \hline
C & 2.3 & 8.8$\pm0.1$  & 36.2$\pm0.6$  & ... & ...\\ 
    & 5.0 & 1.8$\pm0.1$  & 18.7$\pm0.7$ & ... & ... \\
    & 8.4 & 0.77$^\dagger \pm0.09$  & 2.8$\pm0.6$ & ... & ...\\
W & 2.3 & 2.0$\pm0.1$ & 13.6$\pm0.5$ & ... & ...\\ 
    & 5.0 & 0.42$\pm0.09$ & 2.5$\pm0.6$ & ... & ...\\
    & 8.4 & $<0.10^\ddagger$  & $<0.10^\ddagger$ & ... & ...\\
Core$-$East & 15.4 &  0.7$\pm0.2$ & 0.9$\pm0.4$ & (1.0$\times0.2$)$^a$ & (0.5$\times$0.1)\\ 
Core$-$West & 15.4 &  0.8$\pm0.2$ & 0.9$\pm0.4$ & (1.2$\times0.7$)$^a$ & (0.6$\times$0.4) \\    \hline
\end{tabular}   
\\[10pt]
\caption*{Column~1: C is the south-east ``hotspot-jet" structure, W is the north-west ``hotspot-jet" structure. Column~2: Frequency of observation in GHz. Column~3: Peak intensity with error. $\dagger$ = the peak intensity is not from the exact hotspot position but another peak in the same structure. $\ddagger$ The {\it r.m.s.} noise in the image. Column~4: Total flux density {and error}; this is for the entire structure in case of C and W components. Columns~5 \& 6: Major and minor axes of the deconvolved core sizes in milliarcseconds and parsecs, respectively. $^a$ Errors in the convolved core sizes are of the order of 0.5~mas and 0.2~mas for the major and minor axes, respectively.}
\end{table}

\clearpage
\section{Methods}
\subsection{Radio Data Reduction: VLBI}
We have carried out phase-referenced VLBI observations at 2.3, 5, 8 and 15 GHz using 9 antennas of the VLBA on August 28, 2002 (Project ID: BV045). The Brewster antenna could not participate in the experiment due to a power supply failure. The target and the phase calibrator (J2329+0834) were observed in a $2-7-2$ min nodding cycle, with the total time on source ranging from $\approx$1.6 hours at 15~GHz and $\approx$1.4 hours at 2~GHz. We note that the choice of the long nodding cycle was driven by practical constraints (filling a complete pass of the data tape of 44 mins at all the four frequencies), as well as the science goals of the original proposal, which were to obtain good astrometric measurements in order to determine the jet speeds around the bright ``C'' component. The data were reduced following standard procedures in the Astronomical Image Processing System ({\tt AIPS}; see http://www.aips.nrao.edu/CookHTML/CookBookap3.html\#x177-397000C). {Ionospheric dispersive delays were corrected from maps of total electron content using the {\tt AIPS} procedure {\tt VLBATECR}. Images were made directly from the phase-referenced visibilities using the {\tt AIPS} task {\tt IMAGR}; no self-calibration was performed. All images presented in this paper were created using natural weighting with {\tt ROBUST} parameter = 5. The  radio cores at 15~GHz are also detected in images made with uniform weighting using {\tt ROBUST}=$-5$. } The final {\it r.m.s.} noise in the images varies between $\sim1.0\times10^{-4}$~Jy~beam$^{-1}$ at 2~GHz, and $\sim2.0\times10^{-4}$~Jy~beam$^{-1}$ at 15~GHz, respectively. The flux densities, sizes and positions of the compact components were obtained using the {\tt AIPS} Gaussian-fitting task {\tt JMFIT}, while {\tt AIPS} verbs {\tt TVWIN} and {\tt IMSTAT} were used to derive the flux densities of extended components. The major and minor axes corresponding to the size of the core components, as listed in Table~1, are deconvolved from the beam in {\tt JMFIT}. For the western core, the minor axis could not be deconvolved in {\tt JMFIT}; we therefore used the 15~GHz beam minor axis for the core size instead. The spectral index and spectral index noise images were made using the {\tt AIPS} task {\tt COMB}, after first registering the source positions with {\tt OGEOM} (using the brightest, most compact feature, which was the hotspot), the two frequency images made with identical beams (matching the lower resolution image) in {\tt IMAGR}. Flux density values below 3$\sigma$ at both frequencies were blanked before creating the spectral index images. 

\subsection{Radio Data Reduction: EVLA}
We have reduced 15 GHz data from the Expanded Very Large Array (EVLA) in the A-array configuration; these are archival NRAO data from project 14A-471 acquired on March 21, 2014. We calibrated the data using the Common Astronomy Software Applications ({\tt CASA}) package version 4.7.2. pipeline. Imaging and a round of phase self-calibration was also done using {\tt CASA}. The {\it r.m.s.} noise in the final 15~GHz image is $\sim1.0\times10^{-5}$~Jy~beam$^{-1}$ (Figure~\ref{fig1}). A radio core on the $\sim$100~mas scale is detected for the first time in this galaxy. Since no radio core was detected at 1.4~GHz by Momjian et al.\cite{Momjian03}, the core spectral index is inverted; it is $>+0.60$ assuming 4 times the {\it r.m.s.} noise as the upper limit to the core flux density at 1.4~GHz. The beam-deconvolved size of the EVLA core region, as derived using the {\tt JMFIT} task, is 100~mas $\times$ 20~mas. The brightness temperature of the radio core is $1.5\times10^4$~K.

\subsection{Probability of a Noise Peak Mimicking a Radio Core}
In order to check the credibility of the core component(s), we estimated a surface density of noise peaks in the 15~GHz image by dividing the number of noise peaks having a signal-to-noise ratio $\ge5$ by the entire image ($8192~\times~8192$ pixels, with pixel size = 0.13~mas). We found $\approx$105 noise peaks in an image of size $1.065\times1.065$~arcsec$^2$, resulting in a noise peak surface density of $\approx93$ peaks~arcsec$^{-2}$. Considering then a stripe between the C and W hotspots of length equal to the hotspot-hotspot distance (0.515 arcsec) and width equal to the minor axis of the synthesized beam of the 2.3~GHz image showing these hotspots (3.87~mas), $\approx$0.18 noise peaks should be expected in this region. That is, the probability of finding a noise peak in this stripe between the two hotspots is $\sim18\%$; the probability of finding two noise peaks in this stripe is $\sim3$\%. 

We also created several fake u,v-datasets using the AIPS task DTSIM. The frequency setup as well as the gains and  system temperatures were made identical to the real data. A source of flux density 0.9 mJy and RA, DEC offset of $-0.203580, 0.082397$ arcseconds from the phase-centre was included in the fake data. We found that the noise peak distribution was indistinguishable from the real data, implying that systematic noise effects were not dominant in the real image. Similarly, the probability of finding two noise peaks in the stripe described above, varied between $\sim3-6$\% for different datasets with slightly different input noise levels. 

The probability therefore that the core components are {\it two} spurious noise peaks at {\it one third the distance} between the two hotspots is $<0.4\%$. We conclude that the two radio cores cannot be random 5$\sigma$ peaks in the 15~GHz image. 

\subsection{Spectral Index Errors}
We estimated the $8-15$~GHz spectral index limits using four times the {\it r.m.s.} noise in the 8~GHz image ({\it r.m.s.} noise = $1.5\times10^{-4}$~Jy~beam$^{-1}$). Using five times the {\it r.m.s.} noise gives spectral index limits of $+0.03$ and $-0.02$, while using three times the {\it r.m.s.} noise gives spectral index limits of $-0.82$ and $-0.86$. This suggests that the errors in our spectral index limit calculation are of the order of 30\% or larger. However, the fact that the cores are not detected at 8~GHz and lower frequencies, suggests that the overall core spectrum is inverted. 
More data are needed to better constrain the turnover frequency.

The data that support the plots within this paper and other findings of this study are available from the corresponding author upon reasonable request.


\begin{thebibliography}{0}%
\makeatletter
\providecommand \@ifxundefined [1]{%
 \@ifx{#1\undefined}
}%
\providecommand \@ifnum [1]{%
 \ifnum #1\expandafter \@firstoftwo
 \else \expandafter \@secondoftwo
 \fi
}%
\providecommand \@ifx [1]{%
 \ifx #1\expandafter \@firstoftwo
 \else \expandafter \@secondoftwo
 \fi
}%
\providecommand \natexlab [1]{#1}%
\providecommand \enquote  [1]{``#1''}%
\providecommand \bibnamefont  [1]{#1}%
\providecommand \bibfnamefont [1]{#1}%
\providecommand \citenamefont [1]{#1}%
\providecommand \href@noop [0]{\@secondoftwo}%
\providecommand \href [0]{\begingroup \@sanitize@url \@href}%
\providecommand \@href[1]{\@@startlink{#1}\@@href}%
\providecommand \@@href[1]{\endgroup#1\@@endlink}%
\providecommand \@sanitize@url [0]{\catcode `\\12\catcode `\$12\catcode
  `\&12\catcode `\#12\catcode `\^12\catcode `\_12\catcode `\%12\relax}%
\providecommand \@@startlink[1]{}%
\providecommand \@@endlink[0]{}%
\providecommand \url  [0]{\begingroup\@sanitize@url \@url }%
\providecommand \@url [1]{\endgroup\@href {#1}{\urlprefix }}%
\providecommand \urlprefix  [0]{URL }%
\providecommand \Eprint [0]{\href }%
\providecommand \doibase [0]{http://dx.doi.org/}%
\providecommand \selectlanguage [0]{\@gobble}%
\providecommand \bibinfo  [0]{\@secondoftwo}%
\providecommand \bibfield  [0]{\@secondoftwo}%
\providecommand \translation [1]{[#1]}%
\providecommand \BibitemOpen [0]{}%
\providecommand \bibitemStop [0]{}%
\providecommand \bibitemNoStop [0]{.\EOS\space}%
\providecommand \EOS [0]{\spacefactor3000\relax}%
\providecommand \BibitemShut  [1]{\csname bibitem#1\endcsname}%
\let\auto@bib@innerbib\@empty
\end{thebibliography}%


\begin{thebibliography}{41}
\expandafter\ifx\csname natexlab\endcsname\relax\def\natexlab#1{#1}\fi

\bibitem{sanders96} Sanders, D.~B. \& Mirabel, I.~F.  Luminous infrared galaxies. {\araa} \textbf{34}, {749--792} ({1996}).

\bibitem{verdes1997} {{Verdes-Montenegro}, L.} \emph{et~al.} {Hickson 96: a physical compact group.} {\aap} \textbf{321}, {409--423} ({1997}).

\bibitem[{{Lasker} {et~al.}(1990){Lasker}, {Sturch}, {McLean}, {Russell}, {Jenkner}, \& {Shara}}]{Lasker90} {Lasker}, B.~M., {Sturch}, C.~R., {McLean}, B.~J., {Russell}, J.~L., {Jenkner}, H., \& {Shara}, M.~M. {The Guide Star Catalog. I - Astronomical foundations and image processing}, {\aj} \textbf{99}, {2019--2058} (1990).

\bibitem{Miller90} {Miller}, J.~S., \& {Goodrich}, R.~W. Spectropolarimetry of high-polarization Seyfert 2 galaxies and unified Seyfert theories. {\apj} \textbf{355}, 456--467 (1990).

\bibitem{Begelman80} {{Begelman}, M.~C.}, {{Blandford}, R.~D.} \& {{Rees}, M.~J.} {Massive black hole binaries in active galactic nuclei}. {\nat} \textbf{287}, {307--309} ({1980}).

\bibitem{Aguerri01} {Aguerri}, J.~A.~L., {Balcells}, M., \& {Peletier}, R.~F. Growth of galactic bulges by mergers. I. Dense satellites. {\aap} \textbf{367}, 428--442 (2001).

\bibitem{Kormendy13} {Kormendy}, J., \& {Ho}, L.~C. Coevolution (or not) of supermassive black holes and host galaxies. {\araa} \textbf{51}, 511--653 (2013).

\bibitem{Momjian03} {{Momjian}, E.}, {{Romney}, J.~D.}, {{Carilli}, C.~L.} \& {{Troland}, T.~H.} {{Sensitive VLBI continuum and H I absorption observations of NGC 7674: First scientific observations with the combined array VLBA, VLA, and Arecibo}}. {\apj} \textbf{597}, {809--822} ({2003}).

\bibitem{Ulvestad99} {Ulvestad}, J.~S. et~al. Subrelativistic radio jets and parsec-scale absorption in two seyfert galaxies. {\apjl} \textbf{517}, L81--L84 (1999).

\bibitem[{{Barvainis} {et~al.}(1996){Barvainis}, {Lonsdale}, \&
  {Antonucci}}]{Barvainis96}
{Barvainis}, R., {Lonsdale}, C., \& {Antonucci}, R. {\aj} \textbf{111}, {1431--1443} (1996).

\bibitem[{{Kharb} {et~al.}(2014){Kharb}, {O'Dea}, {Baum}, {Hardcastle},
  {Dicken}, {Croston}, {Mingo}, \& {Noel-Storr}}]{Kharb14}
{Kharb}, P., {O'Dea}, C.~P., {Baum}, S.~A., {Hardcastle}, M.~J., {Dicken}, D.,
  {Croston}, J.~H., {Mingo}, B., \& {Noel-Storr}, J.  {\mnras}, {\bf 440}, {2976--2987} (2014).
 
\bibitem[{{Hancock} {et~al.}(2010){Hancock}, {Sadler}, {Mahony}, \& {Ricci}}]{Hancock10} {Hancock}, P.~J., {Sadler}, E.~M., {Mahony}, E.~K., \& {Ricci}, R. {Observations and properties of candidate high-frequency GPS radio sources in the AT20G survey}, {\mnras} {\bf 408}, {1187--1206} (2010).

\bibitem{Ulvestad05} {Ulvestad}, J.~S., {Antonucci}, R.~R.~J., \& {Barvainis}, R. VLBA imaging of central engines in radio-quiet quasars. {\apj} \textbf{621}, {123--129} (2005).

\bibitem[{{Varenius} {et~al.}(2017){Varenius}, {Conway}, {Batejat}, {Mart{\'{\i}}-Vidal}, {P{\'e}rez-Torres}, {Aalto}, {Alberdi}, {Lonsdale}, \& {Diamond}}]{Varenius17} {Varenius}, E., {Conway}, J.~E., {Batejat}, F., {Mart{\'{\i}}-Vidal}, I., {P{\'e}rez-Torres}, M.~A., {Aalto}, S., {Alberdi}, A., {Lonsdale}, C.~J., \& {Diamond}, P. {The population of SNe/SNRs in the starburst galaxy Arp 220. A self-consistent analysis of 20 years of VLBI monitoring}. ArXiv e-prints 1702.04772 (2017).

\bibitem{Perez09} {P{\'e}rez-Torres}, M.~A., {Romero-Ca{\~n}izales}, C., {Alberdi}, A., \& {Polatidis}, A. An extremely prolific supernova factory in the buried nucleus of the starburst galaxy IC 694. {\aap} \textbf{507}, L17--L20 (2009).

\bibitem{Rodriguez06} {Rodriguez}, C. et~al. A compact supermassive binary black hole system. {\apj} \textbf{646}, 49--60 (2006).

\bibitem{Boroson09} {Boroson}, T.~A., \& {Lauer}, T.~R. A candidate sub-parsec supermassive binary black hole system. {\nat} \textbf{458}, 53--55 (2009).

\bibitem[{{Chornock} {et~al.}(2009){Chornock}, {Bloom}, {Cenko}, {Silverman}, {Filippenko}, {Hicks}, {Lawrence}, {Chang}, {Comerford}, {George}, {Modjaz}, {Oishi}, {Quataert}, \& {Strubbe}}]{Chornock09} {Chornock}, R., {Bloom}, J.~S., {Cenko}, S.~B., {Silverman}, J.~M., {Filippenko}, A.~V., {Hicks}, M.~D., {Lawrence}, K.~J., {Chang}, P., {Comerford}, J.~M., {George}, M.~R., {Modjaz}, M., {Oishi}, J.~S., {Quataert}, E., \& {Strubbe}, L.~E. {SDSS J1536+0441: An Extreme ``Double-peaked Emitter,'' Not a Binary Black Hole}, The Astronomer's Telegram, 1955 (2009).

\bibitem[{{Wrobel} \& {Laor}(2009)}]{Wrobel09} {Wrobel}, J.~M., \& {Laor}, A. {Discovery of Radio Emission from the Quasar SDSS J1536+0441, a Candidate Binary Black Hole System}. {\apjl}, {\bf 699}, {L22--L25} (2009).

\bibitem{Woo02} {Woo}, J.-H., \& {Urry}, C.~M. Active galactic nucleus black hole masses and bolometric luminosities. {\apj} \textbf{579}, {530--544} (2002).

\bibitem{Ferrarese00} {{Ferrarese}, L.} \& {{Merritt}, D.} {{A fundamental relation between supermassive black holes and their host galaxies}}.  {\apjl} \textbf{539}, {L9--L12} ({2000}).

\bibitem{Ho08} {Ho}, L.~C. Nuclear activity in nearby galaxies. {\araa} \textbf{46}, 475--539 (2008).

\bibitem[{{Hada} {et~al.}(2011){Hada}, {Doi}, {Kino}, {Nagai}, {Hagiwara}, \& {Kawaguchi}}]{Hada11} {Hada}, K., {Doi}, A., {Kino}, M., {Nagai}, H., {Hagiwara}, Y., \& {Kawaguchi}, N. {An origin of the radio jet in M87 at the location of the central black hole}. {\nat} {\bf 477}, {185--187} (2011).

\bibitem[{{Peterson} \& {Cota}(1988)}]{Peterson88}{Peterson}, B.~M., \& {Cota}, S.~A. {The size of the broad-line region in the Seyfert galaxy NGC 4151}, {\apj} 330, {111--120} (1988).

\bibitem[{{Pozo Nu{\~n}ez} {et~al.}(2015){Pozo Nu{\~n}ez}, {Ramolla}, {Westhues}, {Haas}, {Chini}, {Steenbrugge}, {Barr Dom{\'{\i}}nguez}, {Kaderhandt}, {Hackstein}, {Kollatschny}, {Zetzl}, {Hodapp}, \& {Murphy}}]{Nunez15} {Pozo Nu{\~n}ez}, F., {Ramolla}, M., {Westhues}, C., {Haas}, M., {Chini}, R., {Steenbrugge}, K., {Barr Dom{\'{\i}}nguez}, A., {Kaderhandt}, L., {Hackstein}, M., {Kollatschny}, W., {Zetzl}, M., {Hodapp}, K.~W., \& {Murphy}, M. {The broad-line region and dust torus size of the Seyfert 1 galaxy PGC 50427}, {\aap} {\bf 576}, {A73} (2015).

\bibitem{Merritt13} {{Merritt}, D.} {{Dynamics and evolution of galactic nuclei.}} Princeton University Press ({2013}).

\bibitem{Lal04} {Lal}, D.~V., {Shastri}, P., \& {Gabuzda}, D.~C. Milliarcsec-scale radio structure of a matched sample of Seyfert 1 and Seyfert 2 galaxies. {\aap} \textbf{425}, {99--108} (2004).

\bibitem{Bicknell98} {Bicknell}, G.~V., {Dopita}, M.~A., {Tsvetanov}, Z.~I., \& {Sutherland}, R.~S. Are seyfert narrow-line regions powered by radio jets? {\apj} \textbf{495}, {680--690} (1998).

\bibitem[{{Middelberg} {et~al.}(2004){Middelberg}, {Roy}, {Nagar}, {Krichbaum}, {Norris}, {Wilson}, {Falcke}, {Colbert}, {Witzel}, \& {Fricke}}]{Middelberg04} {Middelberg}, E., {Roy}, A.~L., {Nagar}, N.~M., {Krichbaum}, T.~P., {Norris}, R.~P., {Wilson}, A.~S., {Falcke}, H., {Colbert}, E.~J.~M., {Witzel}, A., \& {Fricke}, K.~J. {Motion and properties of nuclear radio components in Seyfert galaxies seen with VLBI}, {\aap} {\bf 417}, {925--944}, (2004).

\bibitem[{{de Young}(2002)}]{deYoung02} {de Young}, D.~S. {The physics of extragalactic radio sources}.  (2002)

\bibitem{Gopal03} {Gopal-Krishna}, {Biermann}, P.~L., \& {Wiita}, P.~J. The origin of X-shaped radio galaxies: Clues from the Z-symmetric secondary lobes. {\apjl} \textbf{594}, L103--L106 (2003).

\bibitem{Zier05} {Zier}, C. Orientation and size of the `Z' in X-shaped radio galaxies. {\mnras} \textbf{364}, 583--592 (2005).

\bibitem{Merritt02} {{Merritt}, D.} \& {{Ekers}, R.~D.} {{Tracing black hole mergers through radio lobe morphology}}.  {\it Science} \textbf{297}, {1310--1313} ({2002}).

\bibitem{footnote} A role for hydrodynamic forces is invoked also in the so-called ``backflow'' model for X- and Z-shaped sources.\cite{LeahyWilliams84,Worrall95} Gopal-Krishna et al.\cite{Gopal03} and Zier\cite{Zier05} point out that the backflow model contains no obvious mechanism for producing the lateral offset of the secondary lobes in Z-shaped sources, particularly when the orientations of the primary and secondary lobes are roughly orthogonal.

\bibitem{Milosavljevic03} {{Milosavljevi{\'c}}, M.} \& {{Merritt}, D.} {{The final parsec problem}}.  In {{Centrella}, J.~M.} (ed.) {The Astrophysics of Gravitational Wave Sources}, vol. {686} of {American Institute of Physics Conference Series}, {201--210} ({2003}).

\bibitem{Bardeen75} {{Bardeen}, J.~M.} \& {{Petterson}, J.~A.} {{The Lense-Thirring effect and accretion disks around Kerr black holes}}.  {\apjl} \textbf{195}, {L65--L67} ({1975}).

\bibitem{Valtonen07} {{Valtonen}, M.~J.} {{New orbit solutions for the precessing binary black hole model of OJ 287}}.  {\apj} \textbf{659}, {1074--1081} ({2007}).

\bibitem{Graham15} {{Graham}, M.~J.} \emph{et~al.} {{A possible close supermassive black-hole binary in a quasar with optical periodicity}}.  {\nat} \textbf{518}, {74--76} ({2015}).

\bibitem[{{Cordes}(2013)}]{Cordes13}
{Cordes}, J.~M. {Limits to PTA sensitivity: spin stability and arrival time precision of millisecond pulsars} , Classical and Quantum Gravity, 30, {224002} (2013).

\bibitem{Huerta15} {{Huerta}, E.~A.}, {{McWilliams}, S.~T.}, {{Gair}, J.~R.} \& {{Taylor}, S.~R.} {{Detection of eccentric supermassive black hole binaries with pulsar timing arrays: Signal-to-noise ratio calculations}}.  {\prd} \textbf{92}, {063010} ({2015}).

\bibitem{Babak16} {{Babak}, S.} \emph{et~al.} {{European pulsar timing array limits on continuous gravitational waves from individual supermassive black hole binaries}}.  {\mnras} \textbf{455}, {1665--1679} ({2016}).

\bibitem{Rasskazov16} {{Rasskazov}, A.} \& {{Merritt}, D.} {{Evolution of massive black hole binaries in rotating stellar nuclei: Implications for gravitational wave detection}}. arXiv e-prints 1606.07484 ({2016}).

\bibitem[{{Owsianik} \& {Conway}(1998)}]{Owsianik98} {Owsianik}, I., \& {Conway}, J.~E. {First detection of hotspot advance in a Compact Symmetric Object. Evidence for a class of very young extragalactic radio sources}, {\aap} {\bf 337}, {69--79}, (1998).

\bibitem[{{eLISA Consortium} {et~al.}(2013){eLISA Consortium}, {Amaro Seoane},
  {Aoudia}, {Audley}, {Auger}, {Babak}, {Baker}, {Barausse}, {Barke}, {Bassan},
  {Beckmann}, {Benacquista}, {Bender}, {Berti}, {Bin{\'e}truy}, {Bogenstahl},
  {Bonvin}, {Bortoluzzi}, {Brause}, {Brossard}, {Buchman}, {Bykov}, {Camp},
  {Caprini}, {Cavalleri}, {Cerdonio}, {Ciani}, {Colpi}, {Congedo}, {Conklin},
  {Cornish}, {Danzmann}, {de Vine}, {DeBra}, {Dewi Freitag}, {Di Fiore}, {Diaz
  Aguilo}, {Diepholz}, {Dolesi}, {Dotti}, {Fern{\'a}ndez Barranco},
  {Ferraioli}, {Ferroni}, {Finetti}, {Fitzsimons}, {Gair}, {Galeazzi},
  {Garcia}, {Gerberding}, {Gesa}, {Giardini}, {Gibert}, {Grimani}, {Groot},
  {Guzman Cervantes}, {Haiman}, {Halloin}, {Heinzel}, {Hewitson}, {Hogan},
  {Holz}, {Hornstrup}, {Hoyland}, {Hoyle}, {Hueller}, {Hughes}, {Jetzer},
  {Kalogera}, {Karnesis}, {Kilic}, {Killow}, {Klipstein}, {Kochkina},
  {Korsakova}, {Krolak}, {Larson}, {Lieser}, {Littenberg}, {Livas}, {Lloro},
  {Mance}, {Madau}, {Maghami}, {Mahrdt}, {Marsh}, {Mateos}, {Mayer},
  {McClelland}, {McKenzie}, {McWilliams}, {Merkowitz}, {Miller}, {Mitryk},
  {Moerschell}, {Mohanty}, {Monsky}, {Mueller}, {M{\"u}ller}, {Nelemans},
  {Nicolodi}, {Nissanke}, {Nofrarias}, {Numata}, {Ohme}, {Otto},
  {Perreur-Lloyd}, {Petiteau}, {Phinney}, {Plagnol}, {Pollack}, {Porter},
  {Prat}, {Preston}, {Prince}, {Reiche}, {Richstone}, {Robertson}, {Rossi},
  {Rosswog}, {Rubbo}, {Ruiter}, {Sanjuan}, {Sathyaprakash}, {Schlamminger},
  {Schutz}, {Sch{\"u}tze}, {Sesana}, {Shaddock}, {Shah}, {Sheard}, {Sopuerta},
  {Spector}, {Spero}, {Stanga}, {Stebbins}, {Stede}, {Steier}, {Sumner}, {Sun},
  {Sutton}, {Tanaka}, {Tanner}, {Thorpe}, {Tr{\"o}bs}, {Tinto}, {Tu},
  {Vallisneri}, {Vetrugno}, {Vitale}, {Volonteri}, {Wand}, {Wang}, {Wanner},
  {Ward}, {Ware}, {Wass}, {Weber}, {Yu}, {Yunes}, \& {Zweifel}}]{Amaro13}
{eLISA Consortium}, {Amaro Seoane}, P., {Aoudia}, S., {Audley}, H., {Auger},
  G., {Babak}, S., {Baker}, J., {Barausse}, E., {Barke}, S., {Bassan}, M.,
  {Beckmann}, V., {Benacquista}, M., {Bender}, P.~L., {Berti}, E.,
  {Bin{\'e}truy}, P., {Bogenstahl}, J., {Bonvin}, C., {Bortoluzzi}, D.,
  {Brause}, N.~C., {Brossard}, J., {Buchman}, S., {Bykov}, I., {Camp}, J.,
  {Caprini}, C., {Cavalleri}, A., {Cerdonio}, M., {Ciani}, G., {Colpi}, M.,
  {Congedo}, G., {Conklin}, J., {Cornish}, N., {Danzmann}, K., {de Vine}, G.,
  {DeBra}, D., {Dewi Freitag}, M., {Di Fiore}, L., {Diaz Aguilo}, M.,
  {Diepholz}, I., {Dolesi}, R., {Dotti}, M., {Fern{\'a}ndez Barranco}, G.,
  {Ferraioli}, L., {Ferroni}, V., {Finetti}, N., {Fitzsimons}, E., {Gair}, J.,
  {Galeazzi}, F., {Garcia}, A., {Gerberding}, O., {Gesa}, L., {Giardini}, D.,
  {Gibert}, F., {Grimani}, C., {Groot}, P., {Guzman Cervantes}, F., {Haiman},
  Z., {Halloin}, H., {Heinzel}, G., {Hewitson}, M., {Hogan}, C., {Holz}, D.,
  {Hornstrup}, A., {Hoyland}, D., {Hoyle}, C.~D., {Hueller}, M., {Hughes}, S.,
  {Jetzer}, P., {Kalogera}, V., {Karnesis}, N., {Kilic}, M., {Killow}, C.,
  {Klipstein}, W., {Kochkina}, E., {Korsakova}, N., {Krolak}, A., {Larson}, S.,
  {Lieser}, M., {Littenberg}, T., {Livas}, J., {Lloro}, I., {Mance}, D.,
  {Madau}, P., {Maghami}, P., {Mahrdt}, C., {Marsh}, T., {Mateos}, I., {Mayer},
  L., {McClelland}, D., {McKenzie}, K., {McWilliams}, S., {Merkowitz}, S.,
  {Miller}, C., {Mitryk}, S., {Moerschell}, J., {Mohanty}, S., {Monsky}, A.,
  {Mueller}, G., {M{\"u}ller}, V., {Nelemans}, G., {Nicolodi}, D., {Nissanke},
  S., {Nofrarias}, M., {Numata}, K., {Ohme}, F., {Otto}, M., {Perreur-Lloyd},
  M., {Petiteau}, A., {Phinney}, E.~S., {Plagnol}, E., {Pollack}, S., {Porter},
  E., {Prat}, P., {Preston}, A., {Prince}, T., {Reiche}, J., {Richstone}, D.,
  {Robertson}, D., {Rossi}, E.~M., {Rosswog}, S., {Rubbo}, L., {Ruiter}, A.,
  {Sanjuan}, J., {Sathyaprakash}, B.~S., {Schlamminger}, S., {Schutz}, B.,
  {Sch{\"u}tze}, D., {Sesana}, A., {Shaddock}, D., {Shah}, S., {Sheard}, B.,
  {Sopuerta}, C.~F., {Spector}, A., {Spero}, R., {Stanga}, R., {Stebbins}, R.,
  {Stede}, G., {Steier}, F., {Sumner}, T., {Sun}, K.-X., {Sutton}, A.,
  {Tanaka}, T., {Tanner}, D., {Thorpe}, I., {Tr{\"o}bs}, M., {Tinto}, M., {Tu},
  H.-B., {Vallisneri}, M., {Vetrugno}, D., {Vitale}, S., {Volonteri}, M.,
  {Wand}, V., {Wang}, Y., {Wanner}, G., {Ward}, H., {Ware}, B., {Wass}, P.,
  {Weber}, W.~J., {Yu}, Y., {Yunes}, N., \& {Zweifel}, P. {The Gravitational Universe}. ArXiv e-prints 1305.5720 (2013).

\bibitem{LeahyWilliams84} Leahy, J.~P. \& Williams, A.~G. The bridges of classical double radio sources. 
{\mnras} \textbf{210}, {929--951} ({1984}).

\bibitem{Worrall95} Worrall, D.~M., Birkinshaw, M. \& Cameron, R.~A. The X-ray environment of the dumbbell radio galaxy NGC 326.  {\apj}, \textbf{449}, {93--104} ({1995}).
\end{thebibliography}
\end{document}